\documentclass{kluwer}
\usepackage{graphics}
\newdisplay{guess}{Conjecture}

\begin{document}
\begin{article}
\begin{opening}

 \title{Discrete Quantum Gravity: I. Zonal spherical functions of the representations of the SO(4,R) group with respect to the SU(2) subgroup and their  application to the Euclidean invariant weight for the Barrett-Crane model. }

\author{Peter Kramer}
\runningauthor{}
\runningtitle{}
\institute{Institut f\" ur theoretische Physik Universit\" at 
T\" ubingen, 72076 T\" ubingen, Germany}
\author{Miguel \surname{Lorente}} 
\institute{Departamento de F\'{\i}sica, Universidad de Oviedo, 
33007 Oviedo, Spain}

\date{Aril 2, 2008}

\begin{abstract}
Starting from the defining transformations of complex matrices for the $SO(4,R)$ group, we construct the fundamental
representation and the tensor and spinor representations of the group $SO(4,R)$. Given the commutation relations for the 
corresponding algebra, the unitary representations of the group in terms of the generalized Euler angles are
constructed. The crucial step for the Barrett-Crane model in Quantum Gravity is the description of the amplitude for the quantum 4-simplex that is used in the state sum
partition function. We obtain  the zonal spherical functions for the construction of the SO(4,R) invariant weight and associate them to the triangular faces of the 4-simplices.
\end{abstract}

\keywords{$SO(4,R)$ group, tensor representation, spin representation, quantum gra\-vi\-ty, spin networks.}

\end{opening}    

\section{Discrete models in quantum gravity}

The use of discrete models in Physics has become very popular, mainly for two reasons. It helps to find the solutions of
some differential equations by numerical methods, which would not be possible to solve by analytic methods. Besides that,
the introduction of a lattice is equivalent to the introduction of a cut-off in the momentum variable for the field in
order to achieve the finite limit of the solution. In the case of relativistic field equations -like the Dirac,
Klein-Gordon, and the electromagnetic interactions- we have worked 
out some particular cases [1].

There is an other motivation for the discrete models and it is based in some philosophical presuppositions that the
space-time structure is discrete. This is more attractive in the case of general relativity and quantum gravity because it
makes more transparent the connection between the discrete properties of the intrinsic curvature and the background
independent gravitational field.

This last approach was started rigorously by Regge in the early sixties [2]. He introduces some triangulation in a
Riemannian manifold, out of which he constructs local curvature, 
coordinate independent, on the polyhedra. With the help of
the total curvature on the vertices of the discrete manifold he 
constructs a finite action which, in the continuous limit,
becomes the standard Hilbert-Einstein action of general relativity.

Regge himself applied his method (``Regge calculus'') to quantum gravity in three dimensions [3]. In this work he assigns
some representation of the $SU(2)$ group to the edges of the triangles. To be more precise, to every tetrahedron appearing in
the discrete triangulation of the manifold he associates a 6j-symbol in such a way that the spin eigenvalues of the
corresponding representation satisfy sum rules described by the edges and vertices of the tetrahedra. Since the value of
the 6j-symbol has a continuous limit when some edges of the tetrahedra become very large, he could calculate the sum of
this limit for all the 6j-symbols attached to the tetrahedra, and in this way he could compare it with the continuous
Hilbert-Einstein action corresponding to an Euclidean non planar manifold.

A different approach to the discretization of space and time was taken by Penrose [4]. Given some graph representing the
interaction of elementary units satisfying the rules of angular momentum 
without an underlying space, he constructs out of this
network (``spin network'') the properties of total angular momentum as a derived concept. Later this model was applied to
quantum gravity in the sense of Ponzano and Regge. In general, a spin network is a triple $(\gamma, \rho, i)$ where 
$\gamma$ is a graph with a finite set of edges $e$, a finite set of vertices $v$, $\rho_e$ is the representation of a
group G attached to an edge, and $i_v$ is an intertwiner attached to each vertex. If we take the product of the amplitudes
corresponding to all the edges and vertices (given in terms of the representations and intertwiners) we obtain the
particular diagram of some quantum state.

Although the physical consequences of Penrose's ideas were soon considered  to be equivalent to the Ponzano-Regge approach to
quantum gravity [5], the last method was taken as guiding rule in the calculation of partition functions. We can mention a few
results. Turaev and Viro [6] calculated the state sum for a 3d-triangulated manifold with tetrahedra described by 6j-symbols
using the $SU(2)_q$ group. This model was enlarged to 4-dimensional triangulations and was proved by Turaev, Oguri, Crane and
Yetter [7] to be independent of the triangulation (the ``TOCY model'').

A different approach was introduced by Boulatov [8] that led to the same partition function as the TOCY model, but with
the advantage that the terms corresponding to the kinematics and the interaction could be distinguished. For this purpose
he introduced some fields defined over the elements of the groups $SO(3)$, invariant under the action of the group, and
attached to the edges of the tetrahedra. The kine\-ma\-ti\-cal term corresponds to the self interacting field over each edge and
the interaction term corresponds to the fields defined in different edges and coupled among themselves. This method (the
Boulatov matrix model) was very soon enlarged to 4-dimensional triangulations by Ooguri [9]. In both models the fields over
the matrix elements of the group are expanded in terms of the representations of the group and then integrated out, with
the result of a partition function extended to the amplitudes over all tetrahedra, all edges and vertices of the
triangulation.

A more abstract approach was taken by Barrett and Crane, generalizing Penrose's spin networks to 4 dimensions. The novelty
of this model consists in the association of representation of the $SO(4,R)$ group to the faces of the tetrahedra. We will
come back to this model in section 5.

Because we are interested in the physical and mathematical properties of the Barrett-Crane model, we mention briefly some
recent work about this model combined with the matrix model approach of Boulatov and Ooguri [10]. In this work the 2D
quantum space-time emerges as a  Feynman graph, in the manner of the 4d-- matrix models. In this way a spin foam model is
connected to the Feynman diagram of quantum gravity.

In these papers part I and II we try to implement the mathematical consequences of the Barrett-Crane model in both the Euclidean and the Lorentz case, We examine    the
group theory in relation to the triangulation of 4-dimensional manifolds in terms of 4-simplices.

In section 2 and 3 we develop the representation theory for the group SO(4,R) and the algebra so(4), out of which the Biedenharn-Dolginov function
is constructed for the boost transformation. In section 5 we review the Barret-Crane model.
We define the spherical harmonics on a coset space $SO(4,R)/SU(2)^c$, equivalent to the 
sphere $S^3$. The intertwiner of two spherical harmonics yields a zonal spherical function. 
In section 6 we introduce the triple product in
$R^4$ that generalizes the vector product and can be useful for the model. In section 7 we 
apply our results  to the evaluation and interpretation of the state sum for the spin network, which
in the continuous limit  tends to the Hilbert-Einstein action. Using the correspondence between bivectors and generators of SO(4,R) we find a
relation between the area of the triangular faces of the tetrahedra and the spin of the representation.

\section{The groups ${\bf SO(4,R)}$ and ${\bf SU(2) \times SU(2)}$}

The rotation group in 4 dimensions is the group of linear transformations that leaves the quadratic form
$x_1^2+x_2^2+x_3^2+x_4^2$ invariant. The
well known fact that this group is locally isomorphic to $SU(2) \times SU(2)$ enables one to decompose the group action in
the following way:

Take a complex matrix (not necessarily unimodular)
\begin{equation}
\label{I1}
    w=\left( {\matrix{y&z\cr
{-\bar z}&{\bar y}\cr}} \right) \quad ,\quad y=x_1+ix_2,-\bar z=x_3+ix_4,
\end{equation}
where $w$ satisfies $w\,w^+=\det (w)$.

We define the {\it full} group action 
\begin{equation}
\label{I2}
(u_1,u_2):     w\to w'=(u_1)^{-1}wu_2,
\end{equation}
where the inverse $(u_1)^{-1}$ is introduced in order to assure a homomorphic action.
Here $(u_1,u_2)\in SU(2)^L \times SU(2)^R$ generate the left and right action, respectively,
\begin{eqnarray*}
\label{I3}
   u_1&=&\left( {\matrix{\alpha &\beta \cr {-\bar \beta }&{\bar \alpha }\cr}} \right) \in SU(2)^L\;\;,\;\;\alpha \bar \alpha +\beta \bar
\beta =1,\\
u_2&=&\left( {\matrix{\gamma &\delta \cr
{-\bar \delta }&{\bar \gamma }\cr
}} \right) \in SU(2)^R\;\;,\;\;\gamma \bar \gamma +\delta \bar \delta =1.
\end{eqnarray*}
The full group action satisfies:
\begin{equation}
\label{I4}
    w'\,w'^+=\det (w')=w\,w^+=\det (w),
\end{equation}
or $\;\; {x'_1}^2+{x'_2}^2+{x'_3}^2+{x'_4}^2={x_1}^2+{x_2}^2+{x_3}^2+{x_4}^2 \;$, which corresponds to the defining relation for $SO(4,R)$.
More precisely, we have the relation 
\begin{equation}
\label{I5}
 SO(4,R) = SU(2)^L \times SU(2)^R/ Z_2.
\end{equation}
Here $Z_2$ is the matrix group generated by $(-1)$ times  the  $2 \times 2$ identity matrix $e$. Clearly for $u_1=u_2=-e$ the action  eq.(2)
keeps $w$ unchanged.

In order to make a connection with $R^4$, we take only the {\em left} action $w'=u_1w$ and express the matrix elements of
$w$ as a 4-vector
\begin{equation}
\label{I6}
   \left( {\matrix{{y'}\cr
{-\bar z'}\cr
{z'}\cr
{\bar y'}\cr
}} \right)=\left( {\matrix{\alpha &\beta &0&0\cr
{-\bar \beta }&{\bar \alpha }&0&0\cr
0&0&\alpha &\beta \cr
0&0&{-\bar \beta }&{\bar \alpha }\cr
}} \right)\left( {\matrix{y\cr
{-\bar z}\cr
z\cr
{\bar y}\cr
}} \right).
\end{equation}

Substituting $y=x_1+ix_2\;, -\bar z=x_3+ix_4$, and $\alpha =\alpha _1+i\alpha _2\;,\;\beta =\beta _1+i\beta _2$, we get
\begin{equation}
\label{I7}
    \left( {\matrix{{x'_1}\cr
{x'_2}\cr
{x'_3}\cr
{x'_4}\cr
}} \right)=\left( {\matrix{{\alpha _1}&{-\alpha _2}&{\beta _1}&{-\beta _2}\cr
{\alpha _2}&{\alpha _1}&{\beta _2}&{\beta _1}\cr
{-\beta _1}&{-\beta _2}&{\alpha _1}&{\alpha _2}\cr
{\beta _2}&{-\beta _1}&{-\alpha _2}&{\alpha _1}\cr
}} \right)\left( {\matrix{{x_1}\cr
{x_2}\cr
{x_3}\cr
{x_4}\cr
}} \right).
\end{equation}

Obviously, the transformation matrix is orthogonal. Similarly for the right action $w'=wu_2^+$
 we get
\begin{equation}
\label{I8}
    \left( {\matrix{{y'}\cr
{-\bar z'}\cr
{z'}\cr
{\bar y'}\cr
}} \right)=\left( {\matrix{{\bar \gamma }&0&{\bar \delta }&0\cr
0&{\bar \gamma }&0&{\bar \delta }\cr
{-\delta }&0&\gamma &0\cr
0&{-\delta }&0&\gamma \cr
}} \right)\left( {\matrix{y\cr
{-\bar z}\cr
z\cr
{\bar y}\cr
}} \right),
\end{equation}
and after substituting $\gamma =\gamma _1+i\gamma _2\;,\;\delta =\delta _1+i\delta _2$, we get
\begin{equation}
\label{I10}
   \left( {\matrix{{x'_1}\cr
{x'_2}\cr
{x'_3}\cr
{x'_4}\cr
}} \right)=\left( {\matrix{{\gamma _1}&{\gamma _2}&{-\delta _1}&{\delta _2}\cr
{-\gamma _2}&{\gamma _1}&{\delta _2}&{\delta _1}\cr
{\delta _1}&{-\delta _2}&{\gamma _1}&{\gamma _2}\cr
{-\delta _2}&{-\delta _1}&{-\gamma _2}&{\gamma _1}\cr
}} \right)\left( {\matrix{{x_1}\cr
{x_2}\cr
{x_3}\cr
{x_4}\cr
}} \right),
\end{equation}
where the transformation matrix is orthogonal.

If we take the full action
\begin{equation}
 \label{I11}
   \left( {\matrix{{y'}&{z'}\cr
{-\bar z'}&{\bar y'}\cr
}} \right)=\left( {\matrix{\alpha &\beta \cr
{-\bar \beta }&{\bar \alpha }\cr
}} \right)\left( {\matrix{y&z\cr
{-\bar z}&{\bar y}\cr
}} \right)\left( {\matrix{{\bar \gamma }&{-\delta }\cr
{\bar \delta }&\gamma \cr
}} \right),
\end{equation}
we get
\begin{eqnarray}
\label{I12}
  \nonumber \left( {\matrix{{y'}\cr
{-\bar z'}\cr
{z'}\cr
{\bar y'}\cr
}} \right)&=&\left( {\matrix{{\alpha \bar \gamma }&{\beta \bar \gamma }&{\alpha \bar \delta }&{\beta \bar \delta }\cr
{-\bar \beta \bar \gamma }&{\bar \alpha \bar \gamma }&{-\bar \beta \bar \delta }&{\bar \alpha \bar \delta }\cr
{-\alpha \delta }&{-\beta \delta }&{\alpha \gamma }&{\beta \gamma }\cr
{\bar \beta \delta }&{-\bar \alpha \delta }&{-\bar \beta \gamma }&{\bar \alpha \gamma }\cr
}} \right)\left( {\matrix{y\cr
{-\bar z}\cr
z\cr
{\bar y}\cr
}} \right)=\\
&=&\left( {\matrix{\alpha &\beta &0&0\cr
{-\bar \beta }&{\bar \alpha }&0&0\cr
0&0&\alpha &\beta \cr
0&0&{-\bar \beta }&{\bar \alpha }\cr
}} \right)\left( {\matrix{{\bar \gamma }&0&{\bar \delta }&0\cr
0&{\bar \gamma }&0&{\bar \delta }\cr
{-\delta }&0&\gamma &0\cr
0&{-\delta }&0&\gamma \cr
}} \right)\left( {\matrix{y\cr
{-\bar z}\cr
z\cr
{\bar y}\cr
}} \right),
\end{eqnarray}
and taking $y=x_1+ix_2\;,\;-\bar z=x_3+ix_4$ we get the general transformation matrix for the 4-dimensional vector in $R^4$
under the group $SO(4,R)$ as
\begin{equation}
\label{I13}
\left( {\matrix{{x'_1}\cr
{x'_2}\cr
{x'_3}\cr
{x'_4}\cr
}} \right)=\left( {\matrix{{\alpha _1}&{-\alpha _2}&{\beta _1}&{-\beta _2}\cr
{\alpha _2}&{\alpha _1}&{\beta _2}&{\beta _1}\cr
{-\beta _1}&{-\beta _2}&{\alpha _1}&{\alpha _2}\cr
{\beta _2}&{-\beta _1}&{-\alpha _2}&{\alpha _1}\cr
}} \right)\left( {\matrix{{\gamma _1}&{\gamma _2}&{-\delta _1}&{\delta _2}\cr
{-\gamma _2}&{\gamma _1}&{\delta _2}&{\delta _1}\cr
{\delta _1}&{-\delta _2}&{\gamma _1}&{\gamma _2}\cr
{-\delta _2}&{-\delta _1}&{-\gamma _2}&{\gamma _1}\cr
}} \right)\left( {\matrix{{x_1}\cr
{x_2}\cr
{x_3}\cr
{x_4}\cr
}} \right).
\end{equation}

Notice that the eight parameters $\alpha _1,\alpha _2,\beta _1,\beta _2,\gamma _1,\gamma _2,\delta _1,\delta _2$ with the
constraints $\alpha _1^2+\alpha _2^2+\beta _1^2+\beta _2^2=1\;\,,\;\,\gamma _1^2+\gamma _2^2+\delta _1^2+\delta _2^2=1$,
can be considered the Cayley parameters for the $SO(4,R)$ group [11].

\section{Tensor and spinor representations of SO(4,R)}

Given the fundamental 4-dimensional representation of $SO(4,R)$ in terms of the parameters $\alpha ,\beta ,\gamma ,\delta
$, as given in eq. \ref{I13},
\begin{equation}
\label{I14}
x'_\mu =g_{\mu \nu }x_\nu, 
\end{equation}
the tensor representations are defined in the usual way
\begin{eqnarray}
\label{I15}
T_{k'_1k'_2\ldots k'_n}=g_{k'_1k_1}&\ldots& g_{k'_nk_n}T_{k_1k_2\ldots k_n},\\
\nonumber &&\left( {k'_i,k_i=1,2,3,4} \right).
\end{eqnarray}

For the sake of simplicity we take the second rank tensors. We can decompose them into totally symmetric and antisymmetric
tensors, namely,
\vspace{-0,2cm}
\begin{eqnarray*}
\label{I16}
S_{ij}&\equiv& x_iy_j+x_jy_i \qquad \mbox{(totally symmetric)},\\
A_{ij}&\equiv& x_iy_j-x_jy_i \qquad \mbox{(antisymmetric)}.
\end{eqnarray*}
If we substract the trace from $S_{ij}$ we get a tensor that transforms under an irreducible representation. For the
antisymmetric tensor the situation is more delicate. In general we have 
\begin{equation}
\label{I17}
A'_{ij}\equiv x'_iy'_j-x'_jy'_i=\left( {g_{i\ell}g_{jm}-g_{j\ell}g_{im}} \right)A_{\ell m}.
\end{equation}

This representation of dimension 6 is still reducible. For simplicity take the left action of the group given in eq. \ref{I7}. The
linear combination of the antisymmetric tensor components are transformed among themselves in the following way:
\begin{equation}
\label{I18}
\left( {\matrix{{A'_{12}+A'_{34}}\cr
{A'_{31}+A'_{24}}\cr
{A'_{23}+A'_{14}}\cr
}} \right)=\left( {\matrix{{A_{12}+A_{34}}\cr
{A_{31}+A_{24}}\cr
{A_{23}+A_{14}}\cr
}} \right),
\end{equation}
\vspace{-5mm}
\begin{eqnarray}
\label{I19}
\nonumber &&\left( {\matrix{{A'_{12}-A'_{34}}\cr
{A'_{31}-A'_{24}}\cr
{A'_{23}-A'_{14}}\cr
}} \right)=\\
\nonumber &=&\left( {\matrix{{\alpha _1^2+\alpha _2^2-\beta _1^2-\beta _2^2}&{-2\left( {\alpha _1\beta _2-\alpha _2\beta
_1} \right)}&{-2\left( {\alpha _1\beta _1+\alpha _2\beta _2} \right)}\cr {2\left( {\alpha _1\beta _2+\alpha _2\beta _1}
\right)}&{\alpha _1^2-\alpha _2^2+\beta _1^2-\beta _2^2}&{2\left( {\alpha _1\alpha _2-\beta _1\beta _2} \right)}\cr
{2\left( {\alpha _1\beta _1-\alpha _2\beta _2} \right)}&{-2\left( {\alpha _1\alpha _2+\beta _1\beta _2} \right)}&{\alpha
_1^2-\alpha _2^2-\beta _1^2+\beta _2^2}\cr }} \right) \times \\
&\times &\left( {\matrix{{A_{12}-A_{34}}\cr {A_{31}-A_{24}}\cr
{A_{23}-A_{14}}\cr
}} \right).
\end{eqnarray}
In the case of the right action given by eq. \ref{I10} the 6-dimensional representation for the antisymmetrie second rank tensor
decomposes into two irreducible 3-dimensional representation of $SO(4,R)$. 
For this purpose one takes the linear
combination of the components of the antisymmetric tensor as before:
\begin{equation}
\label{I20}
\left( {\matrix{{A'_{23}-A'_{14}}\cr
{A'_{31}-A'_{24}}\cr
{A'_{12}-A'_{34}}\cr
}} \right)=\left( {\matrix{{A_{23}-A_{14}}\cr
{A_{31}-A_{24}}\cr
{A_{12}-A_{34}}\cr
}} \right),
\end{equation}
\vspace{-5mm}
\begin{eqnarray}
\label{I21}
\nonumber &&\left( {\matrix{{A'_{23}+A'_{14}}\cr
{A'_{31}+A'_{24}}\cr
{A'_{12}+A'_{34}}\cr
}} \right)=\\
\nonumber &=&\left( {\matrix{{\gamma _1^2-\gamma _2^2-\delta _1^2+\delta _2^2}&{2\left( {\gamma _1\gamma _2+\delta _1\delta
_2} \right)}&{-2\left( {\gamma _1\delta _1-\gamma _2\delta _2} \right)}\cr {-2\left( {\gamma _1\gamma _2-\delta _1\delta
_2} \right)}&{\gamma _1^2-\gamma _2^2+\delta _1^2-\delta _2^2}&{2\left( {\gamma _1\delta _2+\gamma _2\delta _1}
\right)}\cr {2\left( {\gamma _1\delta _1+\gamma _2\delta _2} \right)}&{-2\left( {\gamma _1\delta _2-\gamma _2\delta _1} \right)}&{\gamma _1^2+\gamma _2^2-\delta _1^2-\delta _2^2}\cr }}
\right) \times \\
&\times &\left( {\matrix{{A_{23}+A_{14}}\cr {A_{31}+A_{24}}\cr
{A_{12}+A_{34}}\cr
}} \right).
\end{eqnarray}
Therefore the 6-dimensional representation for the antisymmetric tensor decomposes into two irreducible 3-dimensional
irreducible representation of the $SO(4,R)$ group.

For the spinor representation of $SU(2)^L$ we take 
\begin{equation}
\label{I22}
\left( {\matrix{{a'_1}\cr
{a'_2}\cr
}} \right)=\left( {\matrix{\alpha &\beta \cr
{-\bar \beta }&{\bar \alpha }\cr
}} \right)\left( {\matrix{{a_1}\cr
{a_2}\cr
}} \right)\;,\quad a_1,a_2\in {\not\subset}
\end{equation}

Let $a^{i_1i_2\ldots i_k}\quad ,\quad \left( {i_1,i_2,\ldots i_k=1,2} \right)$ 
be a set of complex numbers of dimension $2^k$ which
transform under the $SU(2)^L$  group as follows:
\begin{equation}
\label{I23}
a^{i'_1\ldots i'_k}=u_{i'_1i_1}\ldots u_{i'_ki_k}a^{i_1\ldots i_k},
\end{equation}
where $u_{i'_1i_1},u_{i'_2i_2}\ldots $ are the components of $u\in SU(2)^L$. 
If $a^{i_1\ldots i_k}$ is totally symmetric in the indices
$i_1\ldots i_k$ the representation of dimension
$(k+1)$ is irreducible. In an analogous way we can define an irreducible representation of $SU(2)^R$ with respect to the totally symmetric
multispinor of dimension $(\ell+1)$.

For the general group $SO(4,R)\sim SU(2)^L\times SU(2)^R$ we can take a set of totally symmetric multispinors that transform under the
$SO(4,R)$ group as
\begin{equation}
\label{I24}
a^{i'_1\ldots i'_k\,j'_1\ldots j'_\ell }=u_{i'_1i_1}\ldots u_{i'_ki_k}\bar v _{j'_1j_1}
\ldots \bar v _{j'_\ell i_\ell} a^{i_1\ldots i_kj_1\ldots j_\ell }
\end{equation}
where $u_{i'_1i_1}\ldots $ are the components of a general element of $SU(2)^L$ and 
$\bar v _{j'_\ell i_\ell}$ are the components
of a general element of $SU(2)^R$. They define an irreducible representation of $SO(4,R)$ of dimension $(k+1)(\ell +1)$ and with labels
(see next section)
\begin{equation}
\label{I25}
\ell _0={{k-\ell} \over 2}\;\;,\quad \ell _1={{k+\ell} \over 2}+1.
\end{equation}

\section{Representations of the algebra ${\bf so(4,R)}$}

Let $J_1,J_2,J_3$ be the generators corresponding to the rotations in the planes $(x_2,x_3),(x_3,x_1)$, and $(x_1,x_2)$ respectively, and
$K_1,K_2,K_3$ the generators corresponding to the 
rotations (boost) in the planes $(x_1,x_4)$, $(x_2,x_4)$ and $(x_3,x_4)$ respectively. They
satisfy the following conmutation relations:
\begin{eqnarray}
\label{I26}
\nonumber &&\left[ {J_p,J_q} \right]=i\varepsilon _{pqr}J_r\;\quad ,\quad \;p,q,r=1,2,3,\\
\nonumber &&\left[ {J_p,K_q} \right]=i\varepsilon _{pqr}K_r,\\
&&\left[ {K_p,K_q} \right]=i\varepsilon _{pqr}J_r. 
\end{eqnarray}
\medskip 
If one defines  
$\bar A={1 \over 2}\left( {\bar J+\bar K} \right)\quad ,
\quad \bar B={1 \over 2}\left( {\bar J-\bar K} \right),$
		
\noindent with \qquad \qquad   $\bar
J=\left( {J_1,J_2,J_3} \right)\quad ,\quad \bar K=\left( {K_1,K_2,K_3} \right)$, then 
\begin{eqnarray}
\label{I27}
\nonumber && \left[ {A_p,A_q} \right]=i\varepsilon _{pqr}A_r\;
\quad ,\quad \;p,q,r=1,2,3,\\
\nonumber && \left[ {B_p,B_q} \right]=i\varepsilon _{pqr}B_r,\\
&& \left[ {A_p,B_q} \right]=0,
\end{eqnarray}
that is to say, the algebra so(4) decomposes into two simple algebras su(2) + su(2)

Let $\phi _{m_1m_2}$ be a basis where $\bar A^2,A_3$ and $\bar B^2,B_3$ are 
diagonal. Then a unitary irreducible representation for the
sets $\left\{ {A_\pm \equiv A_1 \pm iA_2,A_3} \right\}$ and $\left\{ {B_\pm \equiv B_1 \pm iB_2,B_3} \right\}$ is given by 
\begin{eqnarray}
\label{I28}
\nonumber A_\pm \phi _{m_1m_2}&=&\sqrt {\left( {j_1\mp m_1} \right)\left( {j_1\pm m_1+1} \right)}\phi _{m_1\pm 1,m_2},\\
A_3\phi _{m_1m_2}&=&m_1\phi _{m_1m_2}\quad ,\quad -j_1\le m_1\le j_1,
\end{eqnarray}
\begin{eqnarray*}
\label{I29}
B_\pm \phi _{m_1m_2}&=&\sqrt {\left( {j_2\mp m_2} \right)
\left( {j_2\pm m_2+1} \right)}\phi _{m_1m_2\pm 1},\\
B_3\phi _{m_1m_2}&=&m_2\phi _{m_1m_2}\quad ,\quad -j_2\le m_2\le j_2.
\end{eqnarray*}
We change now to a new basis
\begin{equation}
\label{I30}
\psi _{JM}=\sum\limits_{m_1+m_2=m} \left\langle {{j_1m_1j_2m_2}} \mathrel{\left | {\vphantom {{j_1m_1j_2m_2} {JM}}} \right.
\kern-\nulldelimiterspace} {{JM}} \right\rangle \phi _{m_1m_2}
\end{equation}
that corresponds to the Gelfand-Zetlin basis for so(4), 
\[
\psi _{JM}=\left| {\matrix{{j_1+j_2}&,&{j_1-j_2}\cr
{}&J&{}\cr
{}&M&{}\cr
}} \right\rangle.
\]

In this basis the representation for the generators $\bar J,\bar K$ of so(4) 
are given by [12]
\begin{eqnarray}
\label{I31}
\nonumber J_\pm \psi _{JM}&=&\sqrt {\left( {J\mp M} \right)
\left( {J\pm M+1} \right)}\psi _{JM \pm 1},\\
J_3\psi _{JM}&=&M\psi _{JM},\\
\nonumber K_3\psi _{JM}&=&a_{JM}\psi _{J-1,M}+b_{JM}
\psi _{JM}+a_{J+1,M}\psi _{J+1,M},
\end{eqnarray}
where
 \[a_{JM}\equiv \left( {{{\left( {J^2-M^2} \right)\left( {J^2-\ell _0^2} \right)\left( {\ell _1^2-J^2} \right)} \over {\left(
{2J-1} \right)J^2\left( {2J+1} \right)}}} 
\right)^{{1 \mathord{\left/ {\vphantom {1 2}} \right. \kern-\nulldelimiterspace}
2}},\; \; b_{JM}={{M\ell _0\ell _1} \over {J\left( {J+1} \right)}},\]

\noindent with $\ell _0=j_1-j_2\; \; ,\; \ell _1=j_1+j_2+1$ the labels of the representations.

The representation for $K_1, K_2$ are obtained with the help of the commutation relations.

The Casimir operators are
\begin{equation}
\label{I32}
\left( {\bar J^2+\bar K^2} \right)\psi _{JM}=\left( {\ell _0^2+\ell _1^2-1} \right)\psi _{JM},
\end{equation}
\begin{equation}
\label{I33}
\bar J\cdot \bar K\psi _{JM}=\ell _0\ell _1\psi _{JM}.
\end{equation}
The representations in the bases $\psi _{JM}$ are irreducible in the following cases
\begin{eqnarray*}
\label{I34}
\ell _0&=&j_1-j_2=0,\pm {1 \over 2},\pm 1,\pm {3 \over 2},\pm 2,\ldots,\\
\ell _1&=&j_1+j_2-1=\left| {\ell _0} \right|+1,\left| {\ell _0} \right|+2,\ldots,\\
J&=&\left| {j_1-j_2} \right|,\ldots, j_1+j_2.
\end{eqnarray*}

If we exponentiate the infinitesimal generators we obtain the finite representations of $SO(4,R)$ given in terms of the rotation
angles. An element $U$ of $SO(4,R)$ is given as [13]
\begin{equation}
\label{I35}
U\left( {\varphi ,\theta ,\tau ,\alpha ,\beta ,\gamma } \right)=R_3\left( \varphi  \right)R_2\left( \theta  \right)S_3\left(
\tau  \right)R_3\left( \alpha  \right)R_2\left( \beta  \right)R_3
\left( \gamma  \right),
\end{equation}
where $R_2$ is the rotation matrix in the $(x_1x_3)$ plane, $R_3$ the rotation matrix in the $(x_1x_2)$ plane and $S_3$ the
rotation (``boost'') in the $(x_3x_4)$ plane, and
\[0\le \beta ,\tau ,\theta \le \pi \;\;,\;\;0\le \alpha ,\varphi ,\gamma \le 2\pi.
 \]

In the basis $\psi _{jm}$ the action of $S_3$ is as follows:
\begin{equation}
\label{I36}
S_3\left( \tau  \right)\psi _{jm}=
\sum\limits_{j'} {d_{	J'JM}^{j_1j_2}}\left( \tau  \right)\psi _{J'M},
\end{equation}
where
\begin{equation}
\label{I37}
d_{J'JM}^{(j_1j_2)}\left( \tau  \right)=\sum\limits_{m_1m_2} {\left\langle {{j_1j_2m_1m_2}} \mathrel{\left | {\vphantom
{{j_1j_2m_1m_2} {JM}}} \right. \kern-\nulldelimiterspace} {{JM}} \right\rangle }e^{-i\left( {m_1-m_2} \right)\tau }\left\langle
{{j_1j_2m_1m_2}} \mathrel{\left | {\vphantom {{j_1j_2m_1m_2} {J'M}}} \right. \kern-\nulldelimiterspace} {{J'M}} \right\rangle
\end{equation}
is the Biedenharn-Dolginov function, [14] and [15] IV.3.

From this function the general irreducible representations of the operator $U$ in terms of rotation angles is [13]:
\begin{equation}
\label{I38}
U\left( {\varphi ,\theta ,\tau ,\alpha ,\beta ,\gamma } \right)\psi _{JM}=\sum\limits_{J'M'} {D_{J'M'JM}^{j_1j_2}}\left( {\varphi
,\theta ,\tau ,\alpha ,\beta ,\gamma } \right)\psi _{J'M'},
\end{equation}
where
\begin{equation}
\label{I39}
D_{J'M'JM}^{(j_1j_2)}\left( {\varphi ,\theta ,\tau ,\alpha ,\beta ,\gamma } 
\right)=\sum\limits_{m''} {D_{M'M''}^{J'}\left(
{\varphi ,\theta ,0 } \right)}d_{J'JM''}^{(j_1j_2)}
\left( \tau  \right)D_{M''M}^{J}\left( {\alpha ,\beta ,\gamma } \right).
\end{equation}

We now give some particular values of these representations. In the case of spin $j={1 \mathord{\left/ {\vphantom {1 2}} \right.
\kern-\nulldelimiterspace} 2}$ we know
\begin{equation}
\label{I40}
 R_3\left( \alpha  \right)R_2\left( \beta  \right)R_3\left( \gamma  \right)=\left( {\matrix{{\cos {\beta  \over 2}e^{i{{\alpha
+\gamma } \over 2}}}&{i\sin {\beta  \over 2}e^{-i\left( {{{\gamma -\alpha } \over 2}} \right)}}\cr {i\sin{\beta 
\over 2}e^{i{{\gamma -\alpha } \over 2}}}&{\cos {\beta  \over 2}e^{-i\left( {{{\alpha +\gamma } \over 2}} \right)}}\cr }} \right)
\end{equation}

Introducing the variables
\begin{eqnarray*}
\label{I41}
x_1=\cos {\beta  \over 2}\cos {\alpha + \gamma  \over 2}\;\;,\;\;x_2=\cos {\beta  \over 2}\sin{\alpha + \gamma  \over 2},\\
x_3=\sin {\beta  \over 2}\sin {\gamma - \alpha  \over 2}\;\;,\;\;
x_4=\sin {\beta  \over 2}\cos {\gamma - \alpha  \over 2},
\end{eqnarray*}
we have
\begin{equation}
\label{I42}
R_3\left( \alpha  \right)R_2\left( \beta  \right)R_3\left( \gamma  \right)=\left( {\matrix{{x_1+ix_2}&{x_3+ix_4}\cr
{-x_3+ix_4}&{x_1-ix_2}\cr
}} \right).
\end{equation}
Similarly we have
\begin{equation}
\label{I43}
R_3\left( \varphi  \right)R_2\left( \theta  \right)S_3\left( \tau  \right)=\left( {\matrix{{y_1+iy_2}&{y_3+iy_4}\cr
{-y_3+iy_4}&{y_1-iy_2}\cr
}} \right),
\end{equation}
with
\begin{eqnarray*}
\label{I44}
y_1=\cos {\theta  \over 2}\cos {{\varphi +\tau } \over 2}\;\;,\;\;y_2=\cos {\theta  \over 2}\sin {{\varphi +\tau } \over 2},\\
y_3=\sin {\theta  \over 2}\sin {{\tau -\varphi } \over 2}\;\;,\;\;y_4=\sin {\theta  \over 2}\cos {{\tau -\varphi } \over 2}.
\end{eqnarray*}

For the Biedenharn-Dolginov function we have some particular values, see [15]
IV.2.3,
\begin{eqnarray*}
\label{I45}
&&d_{JMM}^{\left[ {j_+,0} \right]}\left( \tau  \right)=i^{J-M}2^J\sqrt {2J+1}\Gamma \left( {J+1} \right)\times \\
&&\times \left( {{{\Gamma \left(
{M+{3 \over 2}} \right)\Gamma \left( {j_+-M+1} \right)\Gamma \left( {j_+-J+1} \right)\Gamma \left( {J+M+1} \right)} \over {\Gamma
\left( {{3 \over 2}} \right)\Gamma \left( {j_++M+2} \right)\Gamma \left( {j_++J+2} \right)\Gamma \left( {J-M+1} \right)\Gamma
\left( {M+1} \right)}}} \right)^{{1 \over 2}}
\end{eqnarray*}

\begin{equation}
\label{I46}
\times \left( {\sin \tau } \right)^{J-M}C_{j_+-j}^{J+1}\left( {\cos \tau } \right),
\end{equation}
where $j_+\equiv j_1+j_2\;\;,\;\;j_-=j_1-j_2=0$, and $C_n^\nu \left( {\cos \tau } \right)$ are the Gegenbauer (ultraspherical)
polynomials which are related to the Jacobi polynomials by 
\begin{equation}
\label{I47}
C_n^\nu \left( {\cos \tau } \right)={{\Gamma \left( {\nu +{3 \over 2}} \right)\Gamma \left( {2\nu +n} \right)} \over {\Gamma
\left( {2\nu } \right)\Gamma \left( {\nu +n+{1 \over 2}} \right)}}P_n^{\left( {\nu -{1 \over 2},\nu -{1 \over 2}} \right)}\left(
{\cos \tau } \right),
\end{equation}

\section{Relativistic spin network in 4-dimensions}

We address ourselves to the Barrett-Crane model that generalized Penrose's spin networks from three dimensions to four dimensions
[16]. They characterize the geometrical properties of 4-simplices, out of which the tesselation of the 4-dimensional manifold is made,
and then attach to them the representations of $SO(4,R)$.

A geometric 4-simplex $S^4$ in Euclidean space is given by the embedding of an ordered set of 5 points $(0,x,y,z,t)$ in $R^4$ which is
required to be non-degenerate (the points should not lie in any hyperplane). Each triangle in it determines a bivector
constructed out of the vectors for the edges. Barrett and Crane proved that classically, a geometric 4-simplex in Euclidean space
is completely characterized (up to parallel translation and inversion through the origin) by a set of 10 bivectors $b_i$, each
corresponding to a triangle in the 4-simplex and satisfying the following properties:
 \begin{enumerate}
\item [i)]  
the bivector changes sign if the orientation of the triangle is changed;

\item [ii)] each bivector is simple, i.e. is given by the wedge product of two vectors for the edges;

\item [iii)]  if two triangles share a common edge, the sum of the two bivector is simple;

\item [iv)]  the sum (considering orientation) of the 4 bivectors corresponding to the faces of a tetrahedron is zero;

\item [v)]  for six triangles sharing the same vertex, the six corresponding bivectors are linearly independent;

\item [vi)]  the bivectors (thought of as operators) corresponding to triangles meeting at a vertex of a tetrahedron satisfy $|tr b_1\left[
{b_2,b_3}
\right]|>0$, i.e. the tetrahedron has non-zero volume.
 \end{enumerate}
Then Barrett and Crane define the quantum 4-simplex with the help of bivectors (thought as elements of the Lie algebra $SO(4,R)$). They 
associate a representation to each triangle and a tensor to each tetrahedron. The representations chosen should satisfy the
following conditions, corresponding to the geometrical ones:
\begin{enumerate}
\item [i)] different orientations of a triangle correspond to dual representations;

\item [ii)] the representations of the triangle are ``simple'' representations of $SO(4,R)$, i.e. $j_1=j_2$;

\item [iii)] given two triangles, if we decompose the pair of representations of the tetrahedra bounded by it into its Clebsch-Gordan series,
the tensor for the tetrahedron is decomposed into summands which are non-zero only for simple representations;

\item [iv)] the tensor for the tetrahedron is invariant under $SO(4,R)$. 
 \vspace{-0,2cm}
\end{enumerate}

\subsection{Spin foam models and the Barrett-Crane model.}
We revise the geometrical analysis of Barrett and Crane and follow Reisenberger and Rovelli [21] p. 2.  Consider a simplicial complex in $R^4$ and  fix in it a 4-simplex $S^4$. 
This 4-simplex is bounded by five 3-simplices or tetrahedra, by ten 2-simplices or triangles, by ten 1-simplices or edges, and has five vertices.
Any triangle belonging to $S^4$ bounds and determines exactly two tetrahedra of $S^4$, as can be seen by inspection of Fig. 1.

For the dualization of the spin network we follow Reisenberger and Rovelli [21] pp. 2-4 which is in line with the standard dualization of cell complexes [20] pp. 377-382. In the language of [21] the simplicial
complex is denoted as $\Delta$ and its dual 2-skeleton as $J(\Delta)$. 
We denote dual objects by $^*$. 
The dual to the 4-simplex is a vertex $v^*$, the dual
to the five tetrahedra  of $S^4$ are five edges $e^*$, and the duals
to the ten triangles   are ten 2-faces $f^*$. The dual boundaries  corresponding to a fixed 4-simplex $S^4$ all share a single dual vertex $v^*$. 
A dual vertex bounds five dual edges and ten dual faces. 
A single dual edge $e^*$ at a vertex $v^*$ bounds four faces $f^*$.
A single dual face $f^*=f^*_{kl}$ at a dual vertex $v^*$ has  
exactly two bounding dual edges $(e^*_k,e^*_l), k<l, k=1,2,\ldots, 4$ and therefore can be labelled by the pair $(k,l), k<l$.

Following  [21], the coloring of a spin network $\Delta$ is the assignment of 
pairs $c=\{\rho(g), b\}$ to geometric boundaries of $\Delta$, with $\rho(g)$ an irrep of the
chosen group $G$ for an  element $g\in G$, and $b$ intertwiners.
Reisenberger and Rovelli [21] assign the irreps $\rho(g)$ to the
ten dual faces $f^*_{ij}$, and the intertwiners to the edges 
$e^*_l, l=1,\ldots, 5$ of each fixed vertex $v^*(J(\Delta))$. The geometric property that a dual edge at a dual vertex bounds four faces is converted by the coloring into the requirement that the intertwiner for this edge couples the four 
irreps associated to the four faces to an invariant under right action.
Reisenberger and Rovelli [21] p.3  claim that in the TOCY (Turaev-Ooguri-Crane-Yetter) models this intertwiner is reduced 
to the intertwining  of pairs. In their explanation of this pairing
on [21] p. 4 they use twenty instead of ten representations and group elements, labelled in pairs as $(g^i_j, g^j_i), i<j$. 
Their pairwise intertwiner for a fixed face takes the form, [21] eq. (19),
\begin{equation}
\label{p1}
V(g^i_j)= W(g^i_j(g^j_i)^{-1}). 
\end{equation}
We shall show in part II section 4 for general groups
that a group representation depending on $g_1(g_2)^{-1}$  like in eq. \ref{p1} arises from  the intertwing
of two irreps to an invariant under the right action 
$(g_1,g_2) \rightarrow (g_1q, g_2q), q \in G$. For  the group
$SO(4,R)$ we get this function in terms of the Gelfand-Zetlin 
representation eq. \ref{g2} in the bracket notation,
\begin{equation}
 \label{p2}
\langle (j_1j_2) J'M'|T_{g_1}T_{(g_2)^{-1}}|(j_1j_2), JM\rangle
\end{equation}

Unfortunately the
doubling of the number of irreps and group elements proposed in [21] and their
pairing has no natural counterpart in the geometry of the spin network. If we modify the
coloring of $J(\Delta)$ such that irreps are attached to dual 
edges and intertwiners to dual faces at a dual vertex, the representations and group elements would pair naturally, and the intertwiners
would couple the five irreps, functions of five group elements, in ten pairs in a form as 
in eq. \ref{p1}. This modified coloring would naturally represent the geometric property
that any pair of dual edges bounds exactly one dual face. 

A second observation  arises from the  use of the 
trace of representations in [21]. If from eq. \ref{p2}
we take the trace of the representation, we obtain
\begin{equation}
\label{t1}
{\rm Trace}( D^{j_1j_2}(g_1(g_2)^{-1})) = \chi^{j_1j_2} (g_1(g_2)^{-1}),
\end{equation}
that is, the character $\chi^{j_1j_2}$ of the irrep. It is easy to see that 
this expression now is invariant not only under right action but also
under the left action $(g_1,g_2) \rightarrow (qg_1, qg_2), q \in G$.
A weaker alternative to this trace formation leads to zonal spherical functions, as we explain in  the next
subsection.

\subsection{Spherical harmonics, simple representations and spherical functions.} 

The spherical harmonics are functions on a 
coset or quotient space $SO(4,R)/SU(2)^c\sim S^3$. We shall derive the spherical harmonics from particular representations on a coset space by the condition that they be left-invariant under $SU(2)^c$.
To determine the stability group 
consider in eq. \ref{I1}
the point $P_0: (x_1,x_2,x_3,x_4)=(1,0,0,0)$ of the sphere $S^3 \in R^4$. In the matrix notation eq. \ref{I1}, the point $P_0$  corresponds to the unit matrix $w_0=e$.
With respect to the actions eq. \ref{I2}, this point is stable under
any action $w_0 \rightarrow u^{-1} w_0u$. These elements form a subgroup 
$SU(2)^c<SO(4,R)$ equivalent to $SU(2)$ with elements $(v_1,v_1)$.
The corresponding coset space $SO(4,R)/SU(2)^c$ can be parametrized
by choosing in eq. \ref{I1} $w=u'\in SU(2)^R$, see eq. \ref{g4} below.

For the present purpose we use the Gelfand-Zetlin 
irrep of $SO(4,R)$ as constructed in section 4. 
We write  these  
irreps for  $(u_1,u_2)\in SO(4,R)$ in a bracket notation 
\begin{eqnarray}
\label{g2}
&&\langle (j_1j_2) J'M'|T_{(u_1,u_2)}|(j_1j_2), JM\rangle
\\ \nonumber
&& := \sum_{m_1'm_2'm_1m_2}
\langle j_1m_1'j_2m_2'|J'M'\rangle
\\ \nonumber 
&&
D^{j_1}_{m_1'm_1}(u_1) D^{j_2}_{m_2'm_2}(u_2)
\langle j_1m_1j_2m_2|JM\rangle. 
\end{eqnarray}
Consider now the restriction of the irrep eq. \ref{g2} to the action of the subgroup $SU(2)^c$ with elements
$(u_1,u_2)\rightarrow  (v_1,v_1)$ .   
We obtain
\begin{equation}
\label{g3}
\langle (j_1j_2) J'M'|T_{(v_1,v_1)}|(j_1j_2), JM\rangle
= \delta_{J'J} D^j_{M'M}(v_1) 
\end{equation}
In other words, the Gelfand-Zetlin basis is explicitly reduced with respect to the stability subgroup $SU(2)^c$.  Next we rewrite a general element 
of $SO(4,R)$ in the form 
\begin{equation}
\label{g4}
(u_1,u_2)=(v_1,v_1)(e,v_2)=(v_1, v_1v_2),\: v_2\in SU(2)^R
\end{equation}
These equations show that the cosets of the stability group
$SU(2)^c < SO(4,R)$ are in one-to-one correspondence to the elements 
$(e,v_2)$ of the subgroup $SU(2)^R < SO(4,R)$ of eqs. \ref{I2}, \ref{I3}. 

Evaluation in the new basis yields in particular
\begin{eqnarray}
\label{g5}
&&\langle (j_1j_2) J'M'|T_{(e,u_2)}|(j_1j_2), JM\rangle
\\ \nonumber 
&&=\sum_{m_1'm_1m_2'm_2}
\delta_{m_1'm_1} D^{j_2}_{m_2'm_2}(v_2)
\\ \nonumber
&&\langle j_1m_1'j_2m_2'|J'M'\rangle
\langle j_1m_1j_2m_2|JM\rangle. 
\end{eqnarray}
It follows that the full representation under restriction to $SU(2)^R$
is given in terms of the irrep $D^{j_2}(v_2)$
of $SU(2)^R$. If we choose in eq. \ref{g5} $(j'm')=(00)$, we assure
from eq. \ref{g3} that all the matrix elements 
\begin{eqnarray}
\label{g6}
&&\langle (j_1j_2)00|T_{(e,v_2)}|(j_1j_2) JM\rangle
\\ \nonumber
&&=\delta_{j_1j_2}\sum_{m_1'm_1m_2'm_2}
\delta_{m_1'm_1} D^{j_2}_{m_2'm_2}(v_2)
\\ \nonumber
&&\langle j_2m_1'j_2m_2'|00\rangle
\langle j_2m_1j_2m_2|JM\rangle
\\ \nonumber 
&&=\delta_{j_1j_2}
\sum_{m_1'm_1m_2'm_2}
(-1)^{(j_2-m_1')}\delta_{m_1'm_1} D^{j_2}_{m_2'm_2}(v_2)
\\ \nonumber
&&\delta_{m_1',-m_2'}\frac{1}{\sqrt{2j_2+1}}
\langle j_2m_1j_2m_2|JM\rangle
\\ \nonumber
&& =\delta_{j_1j_2}\frac{1}{\sqrt{2j_2+1}}
\sum_{m_1m_2}
(-1)^{(j_2-m_1)} D^{j_2}_{-m_1m_2}(v_2)
\langle j_2m_1j_2m_2|JM\rangle
\end{eqnarray}
are invariant under left action with elements $(v_2,v_2) \in  SU(2)^c$.
By definition these  are the spherical harmonics 
on $SO(4,R)/SU(2)^c$.

We summarize  these results for  spherical harmonics on 
$SO(4,R)/SU(2)^c$ in

\noindent
{\bf 1  Theorem: Spherical harmonics of $SO(4,R)$}:\\
 (a) {\em Domain}: The spherical harmonics are defined on the coset space for the stability group  $SU(2)^c$ of
the sphere $S^3$. This coset space  
from eq. \ref{g4} can be taken in the form $SU(2)^R$.\\
(b) {\em Characterization}: The spherical harmonics on this coset space are given 
by the matrix elements eq. \ref{g6} of simple irreps.\\
(c) {\em Transformation properties}: Under right action of $SO(4,R)$, the  spherical harmonics eq. \ref{g6}
transform according to simple irreps $D^{j_2j_2}$, which in the Gelfand Zetlin basis are given by eq. \ref{g6} with $j_1=j_2$. Any   left action by $(v_1,v_1)\in SU(2)^c$
leaves the expressions eq. \ref{g6}, taken as matrix elements of the 
full irrep, invariant.\\
(d) {\em Measure}: The spherical harmonics
form a complete  orthonormal set on the coset space $SO(4,R)/SU(2)^c$.
The measure on $SO(4,R)$ from eq. \ref{I4} is the product of two measures 
for groups $SU(2)$. It follows that the measure on the coset space 
$SO(4,R)/SU^c(2)$ has the form of a measure $d\mu(u)$ on $SU(2)^R$.

The coloring of the spin network in [21] attaches irreps $\rho(g)$ 
of the group $G$ and intertwiners to geometric boundaries. For given group element
$g \in G$, the full representation is fixed by an irrep label $\lambda$ and
sets of row and column labels.  This coloring scheme can easily be modified 
by attaching only subsets of matrix elements to a geometric boundary.
The  particular choice of matrix elements eq. \ref{g6} implies that  spherical harmonics are attached. The use of a coset space $SO(4,R)/SU(2)^c$ and  of functions on these  
for spin networks is advocated 
by Freidel et al. [22] pp. 14-16. We agree with these authors but strictly distinguish between 
spherical harmonics and simple representations which determine their
transformation properties. Spherical harmonics by eq. \ref{g6}
are particular matrix elements of simple irreps and live on the coset space
$SO(4,R)/SU(2)$, not on the full group space of $SO(4,R)$.

The results of Theorem 1 allow us to comment on the Kronecker product of
simple irreps of $SO(4,R)$. Reisenberger and Rovelli [21] p. 3 noted correctly that the
Kronecker product of two simple irreps of $SO(4,R)$ contain both simple 
and non-simple irreps. To avoid the non-simple ones they introduce 
projectors. If we replace simple irreps by the spherical harmonics 
of eq. \ref{g6}, the situation changes. A product of two spherical harmonics 
is still a function on the same coset space $SO(4,R)/SU(2)^c$.
Since the spherical harmonics form a complete set on this coset space,
such a product can be expanded again exclusively in spherical harmonics.
The expansion coefficients are those particular coupling coefficients 
which relate products of simple irreps to simple irreps. The non-simple 
irreps automatically drop out of these expansions. 
The measure on the coset space from Theorem 1(d) is only a factor of the measure on the full
group space and equivalent to the measure on the single group $SU(2)^R$.

\subsection{Spherical harmonics and zonal spherical functions.}

The intertwiners appearing in the spin networks correspond to 
right-hand coupling of pairs of irreps to a function invariant under right action, see part II section 4.
This right-hand coupling applies as well to the coupling of pairs of
spherical harmonics. Taking these in the form of eq. \ref{g6} as functions of
group elements $(g_1,g_2)$ yields the expression
\begin{equation}
\label{g7}
f^{(j_2j_2)}(g_1(g_2)^{-1}):= \langle {j_2j_2} 00 |g_1(g_2)^{-1}|{j_2j_2} 00\rangle.
\end{equation}
The function of $g_1(g_2)^{-1}$ on the righthand side of eq. \ref{g7} is
a zonal spherical function on $SO(4,R)$ with respect to the subgroup
$SU(2)^c$. For a general group $G$ with subgroup $SU(2)$ we refer to part II section 8. 
A  zonal spherical 
is a matrix elements of an irrep $D^{\lambda}(g) $ characterized by 
the invariance  both under left- and right-action with $h \in H$. 
The Gelfand-Zetlin irrep eq. \ref{g2} of $SO(4,R)$ is adapted to the subgroup 
$SU(2)^c < SO(4,R)$ with subgroup representation labels  $(J'M'), (JM)$.

{\bf 2 Def}: A zonal spherical function of $g=(u_1,u_2)$ for 
the subgroup $SU(2)^c < SO(4,R)$ is  
given in the Gelfand-Zetlin basis eq. \ref{g2} by 
\begin{equation}
 \label{g7a}
f^{(j_2j_2)}(u_1,u_2):=\langle {j_2j_2} 00 |T_{(u_1,u_2)}|{j_2j_2} 00\rangle. 
\end{equation}

The expression eq. \ref{g7a},
in contrast to the trace eq. \ref{t1}, is not invariant under 
general left actions. It has the weaker invariance
\begin{equation}
\label{g8}
f^{(j_2j_2)}((h_1)^{-1}gh_2) = f^{(j_2j_2)}(g),\; (h_1, h_2) \in SU(2)^c.
\end{equation}
and so it lives on the double cosets of $SO(4,R)$ with respect to
$SU(2)^c$. The zonal spherical functions eq. \ref{g7a} must be distinguished from the spherical functions discussed by Godement in [18].

By use of the angular parameters introduced in eq. \ref{I35}, we 
obtain the zonal spherical function eq. \ref{g7a} in terms of the single parameter 
$\tau$:
\begin{eqnarray}
\label{g8a}
&&f^{(j_2j_2)} (\tau )
\\ \nonumber 
&& =\sum_{m_1  + m_2  = 0} \langle j_2 m_1 j_2 m_2 | 00  \rangle 
\exp(- i(m_1  - m_2 )\tau)  \langle j_2 m_1 j_2 m_2 | 00 \rangle
\\ \nonumber
&&=\frac{1}{2j_2+1} \sum_{m_1  =  - j_1 }^{j_1 } \exp( -i2m_1\tau)   = \frac{1}{2j_2+1}\frac{\sin((2j_1  + 1)\tau)}{\sin \tau}
\end{eqnarray}
{\bf 3 Theorem}: The zonal spherical functions for simple irreps of $SO(4,R)$ with subgroup 
$SU(2)^c$ given by eq. \ref{g7a} become the  functions eq. \ref{g8a} of the  parameter $\tau$.

Pairs of spherical harmonics can still be intertwined to invariants 
under the right action of SO(4,R). The result of this intertwining 
is a zonal spherical function of the type eq. \ref{g7a} of
the product $(g_1(g_2)^{-1}$ of two group elements and by eq. \ref{g8a}
can be given as a function of the angular parameter $\tau$ for the 
group element $g_1(g_2)^{-1}$.

\noindent{\bf 4 Theorem}: If, in agreement with [22], not full simple irreps but
spherical harmonics are attached to  boundaries of the spin network, any pairwise 
intertwiner becomes a zonal spherical function  
$f^{(j_2j_2)}(g_1(g_2)^{-1})$ eq. \ref{g7a}.

\section{The triple product in $R^4$}

Before we apply the representation theory developed in previous sections to the Barrett-Crane model we introduce some geometrical
properties based in the triple product that generalizes the vector (cross) product in $R^3$. Given three vectors in $R^4$, we
define the triple product:
\begin{eqnarray}
\label{g10}
\nonumber u\wedge v\wedge w&=&-v\wedge u\wedge w=-u\wedge w\wedge v=-w\wedge v\wedge u=v\wedge w\wedge u=\\
\nonumber &=&w\wedge u\wedge v,\\
u\wedge u\wedge v&=&u\wedge v\wedge u=v\wedge u\wedge u=0.
\end{eqnarray}

If the vectors in $R^4$ have cartesian coordinates \[u=\left( {u_1,u_2,u_3,u_4} \right)\;,\;v=\left( {v_1,v_2,v_3,v_4}
\right)\;,\;w=\left( {w_1,w_2,w_3,w_4} \right),\] we define an orthonormal basis in $R^4$
\[ \hat \imath=\left( {1,0,0,0} \right)\;,\;\;\hat \jmath=\left( {0,1,0,0} \right)\;,\;\;\hat k=\left( {0,0,1,0}
\right)\;,\;\;\hat
\ell =\left( {0,0,0,1} \right).
\]
The triple product of these vectors satisfies

\[\hat \imath\wedge \hat \jmath\wedge \hat k=-\hat \ell \;,\;\;\hat \jmath\wedge \hat k\wedge \hat \ell =\hat \imath\;,\;\;\hat
k\wedge \hat \ell \wedge \hat \imath=-\hat \jmath\;,\;\;\hat \imath\wedge \hat \jmath\wedge \hat \ell =\hat k\;.\]

In coordinates the triple product is given by the determinant
\begin{equation}
\label{g11}
u\wedge v\wedge w=\left| {\matrix{{\hat \imath}&{\hat \jmath}&{\hat k}&{\hat \ell }\cr
{u_1}&{u_2}&{u_3}&{u_4}\cr
{v_1}&{v_2}&{v_3}&{v_4}\cr
{w_1}&{w_2}&{w_3}&{w_4}\cr
}} \right|.
\end{equation}

The scalar quadruple product is defined by
\begin{eqnarray}
\label{g12}
\nonumber a\cdot \left( {b\wedge c\wedge d} \right)&=&\left| {\matrix{{a_1}&{a_2}&{a_3}&{a_4}\cr
{b_1}&{b_2}&{b_3}&{b_4}\cr
{c_1}&{c_2}&{c_3}&{c_4}\cr
{d_1}&{d_2}&{d_3}&{d_4}\cr
}} \right|=\left[ {abcd} \right]=-\left[ {abdc} \right]=\\ 
&=&-\left[ {acbd} \right]=\left[ {acdb} \right] \mbox {and so on}.
\end{eqnarray}
It follows: $a\cdot a\wedge b\wedge c=b\cdot a\wedge b\wedge c=c\cdot a\wedge b\wedge c=0$.

We can use the properties of the three vector for the description of the 4-simplex. Let $\{ 0,x,y,z,t\}$ be the 4-simplex in
$R^4$. Two tetrahedra have a common face

$\left\{ {0,x,y,z} \right\}\cap \left\{ {0,x,y,t} \right\}=\left\{ {0,x,y} \right\}.$

Each tetrahedron is embedded in an hyperplane characterized by a vector perpendicular to all the vectors forming the tetrahedron.
For instance,
 
$\{ 0,x,y,z\}$ is characterized by $a=x\wedge y\wedge z,$

$\{ 0,x,y,t\}$ is characterized by $b=x\wedge y\wedge t.$
\begin{center} 
\includegraphics{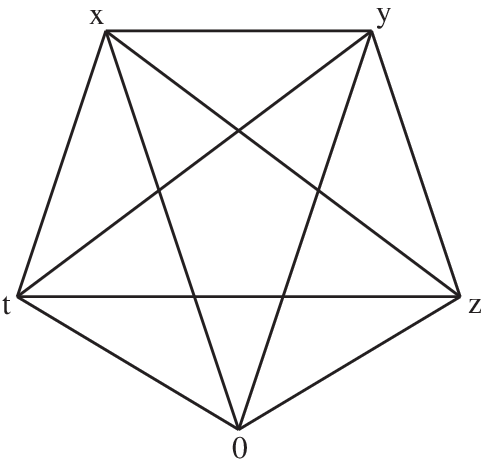}
\end{center}

Fig. 1. A simplex $S^4$ in $R^4$ seen in a projection to 
a two dimensional plane.

The vector $a$ satisfies $a\cdot x=a\cdot y=a\cdot z=0$, 

the vector $b$ satisfies $b\cdot x=b\cdot y=b\cdot t=0.$

The triangle $\{0,x,y\}$ shared by the two tetrahedra is characterized by the bivector $x\wedge y$. The plane where the triangle
is embedded is defined by the two vectors $a,b$, forming the angle $\phi $, given by 
\[ \cos \phi =a\cdot b. \]
The bivector $a\wedge b$ can be calculated with the help of bivectors  $x\wedge y$, namely,
\[ a\wedge b=\left[ {x\,y\,z\,t} \right]\;^*\left( {x\wedge y} \right).\]
Obviously $a\wedge b$ is perpendicular to $x\wedge y$ 
\begin{equation}
\label{g13}
\left\langle {a\wedge b,x\wedge y} \right\rangle =\left( {a\cdot x} \right)\left( {b\cdot y} \right)-\left( {a\cdot y}
\right)\left( {b\cdot x} \right)=0.
\end{equation}
For completeness we add some useful properties of bivectors in $R^4$. The six components of a bivector can be written as 
$$\begin{array}{lll} 
B_{\mu\nu }=x_\mu y_\nu -x_\nu y_\mu \;\quad ,\quad \;&\mu ,\nu =1,2,3,4\;\quad ,\quad \; &B=\left( {\bar J,\bar K} \right),\\
J_1=\left( {x_2y_3-x_3y_2} \right)\;\quad ,\quad \;&J_2=\left( {x_3y_1-x_1y_3}
\right)\;\quad ,\quad \;&J_3=\left( {x_1y_2-x_2y_1} \right),\\
K_1=\left( {x_1y_4-x_4y_1} \right)\quad ,\quad &K_2=\left( {x_2y_4-x_3y_1} \right)\quad ,\quad &K_3=\left(
{x_3y_4-x_4y_1} \right).
\end{array}$$
The six components of the dual of a bivector are 

\noindent $^*B_{\alpha \beta }={1 \over 2}B_{\mu \nu }\,\varepsilon _{\mu \nu \alpha
\beta}  \;,\quad ^*B=\left( {\bar K,\bar J} \right).$

We take the linear combinations of $\bar J,\bar K$
\begin{equation}
\label{g14}
\bar M={1 \over 2}\left( {\bar J+\bar K} 
\right),\quad \quad \bar N={1 \over 2}\left( {\bar J-\bar K} \right).
\end{equation}
 They form the bivector $\left( \bar M,\bar N \right)$, whose dual is:
\begin{equation}
\label{g15}
^*\left( {M,N} \right)=\left( {M,-N} \right),
\end{equation}
therefore $\bar M$ can be considered the self-dual part, $\bar N$ the antiselfdual part of the bivector $(\bar M,\bar N)$.
$\bar M$ and $\bar N$ coincides with the basis for the irreducible tensor representations of section 3. The norm of the
bivectors can be explicitly calculated.
\begin{eqnarray}
\label{g16}
\nonumber\left\| B \right\|^2&=&\left\langle {B,B} \right\rangle =J^2+K^2=\left\| x \right\|^2\left\| y \right\|^2-\left| {x,y}
\right|^2=\\
&=&\left\| x \right\|^2\left\| y \right\|^2\sin ^2\phi 
(x,y)=4(area)^2\left\{ {0,x,y} \right\},
\end{eqnarray}
\begin{equation}
\label{g17}
\left\| {^*B} \right\|^2=\left\langle {^*B,^*B} \right\rangle =J^2+K^2=\left\| B \right\|^2.
\end{equation}
Finally, the scalar product of two vectors in $R^4$ can be expressed in terms of the corresponding $SU(2)$ matrices

Let $X\Leftrightarrow \left( {\matrix{{x_1+ix_2}&{x_3+ix_4}\cr
{-x_3+ix_4}&{x_1-ix_2}\cr
}} \right),\quad \quad Y\Leftrightarrow 
\left( {\matrix{{y_1+iy_2}&{y_3+iy_4}\cr
{-y_3+iy_4}&{y_1-iy_2}\cr
}} \right).$

Then  
\begin{equation}
\label{g18}
{1 \over 2} Tr\left( {XY^+} \right)=x_1y_1+x_2y_2+x_3y_3+x_4y_4.
\end{equation}

\section{Evaluation of the state sum for the 4-dimensional spin network}

We adopt the geometry of the spin network as explained at the end of section 5.1
In order to evaluate the state sum for a particular triangulation of the total $R^4$ space by 4-simplices, we assign an element $g_k  \in SO(4,R)$
and representation $\rho_k(g_k)$ to each tetrahedron ($k = 1,2,3,4,5)$ of $S^4$ and an intertwiner of $SO(4,R)$  to each triangle of $S^4$ shared by two
tetrahedra. From this triangulation we obtain a dual 2-complex  where 
two dual edge  correspond to the two tetrahedra and a dual face to the triangle, with the two edges  bounding  the dual face. Dually  we attach the representations  $\rho_k (g_k)$ and $\rho_l (g_l )$ of $SO(4,R)$ to the edges $e^*_k$ and $e^*_l$ and contract both representations at the dual face $f^*_{kl}$, giving 
\begin{equation}
\label{g18a}
f^{(j_2j_2)}(g_k g_l^{ - 1} ).
\end{equation}
Here $f^{(j_2j_2)}$ is the contraction of the two simple representations of $SO(4,R)$ to an invariant under right action, compare section 5.3 and  part II section 4.
This contraction is shown to require that the two representations 
of $SO(4,R)$ be equivalent, $\rho_k \sim \rho_l \sim \rho_{kl}$.
Since each element $g \in SO(4,R)$ is a pair $(u_1,u_2)$ of elements of $SU(2)$
and the representations are simple,
the expression eq. \ref{g7a} reduces to a product of two expressions in terms of
$SU(2)$ with the same representation $j_2$. The expression eq. \ref{g7a} has one more implication which we pointed out in section 5.3:  It is valid only 
if the irreps $\rho(g)$ attached  to the tetrahedra are  replaced  by the spherical harmonics eq. \ref{g6}.  Then the intertwiner of a pair of spherical harmonics 
becomes a zonal spherical function eq. \ref{g7a}.

The state sum for the 2-dimensional complex (the Feymann graph of the model) is obtained by taking the product
expression eq. \ref{g18a} for all the edges of the graph and integrating over all the copies of elements of $SO(4,R)$. 
 Barrett and Crane  construct a state sum  $Z_{BC}$ for the quantum
4-simplex in terms of amplitudes $A$, functions  of the colorings and intertwiners attached to simplices of the spin network:

\begin{equation}
\label{g19}
Z_{BC}=\sum\limits_J {\prod\limits_{{\rm triang.}} {A_{{\rm tr}}}}\prod\limits_{{\rm tetrahedra}} {A_{{\rm
tetr.}}}\prod\limits_{4-{\rm simplices}} {A_{{\rm simp.}}}
\end{equation}
where the sum extends to all possible values of the representations $J$.
All the amplitudes $A$ can be expressed by intertwiners of 
pairs of irreps and group elements, and by corresponding zonal spherical functions.
Due to the properties of zonal spherical functions,  the expression 
eq. \ref{g19} is in addition invariant under left and right multiplication with arbitrary  elements of $SU(2)^c<SO(4,R)$.
We can obtain a particular value of eq. \ref{g8a} for $j_2 = {1 \mathord{\left/ {\vphantom {1 2}} \right. \kern-\nulldelimiterspace} 2}$ if we take the
elements $g_k$ and $g_l$ as pairs of unit vectors in $R_4$, say, $x$ and $y$, and use eqs. \ref{I42}, \ref{I43} to obtain
\begin{equation}
 \label{g20}
 f^{(\frac{1}{2}\frac{1}{2})}( xy^+) = x \cdot y = \cos(\varphi)
\end{equation}
where $\varphi $ is the angle between the vectors $x$ and $y$.

The two vectors $(x,y)$ are perpendicular to the hyperplanes where the tetrahedra $k$ and $l$ are embedded, and correspond to the vectors
perpendicular to the face shared by the two tetrahedra, as explained in [17].

With eq. \ref{g7a} it is still possible to give a geometrical interpretation of the probability amplitude encompassed in the zonal spherical function. In fact, the
spin dependent factor appearing in the exponential of eq. \ref{g8a}, 
\begin{equation}
e^{i\left( {2j_{k\ell }+1} \right)\tau _{k\ell }},
\end{equation}
corresponding to two tetrahedra $k, \ell$ intersecting in the triangle $k\ell$, can be interpreted as the product of the angle
between the two vectors $g_k, g_{\ell}$ perpendicular to the triangle and the area $A_{k\ell}$ of the intersecting triangle.

For the proof we identify the component of the antisymmetrie tensor $\left( {\bar J,\bar K} \right)$ with the components of the
infinitesimal generators of the $SO(4,R)$ group
\[J_{\mu \nu }\equiv i\left( {x_\mu {\partial  \over {\partial x_\nu }}-x_\nu {\partial  \over {\partial x_\mu }}} \right).
\]
From eq. \ref{g16}  and eq. \ref{g17} we have
$\left\| {B} \right\|^2=4\left( {A_{k\ell }} \right)^2=2\left( {\bar M^2+\bar N^2} \right)$

But $\bar M^2$ and $\bar N^2$ are the Casimir operators of the $SU(2) \times SU(2)$ group with eigenvalues $j_1\left( {j_1+1}
\right)$ and $j_2\left( {j_2+1}\right)$.

For large values of $j_1=j_2=j_{k \ell}$ we have
\begin{equation}
2\left( {\bar M^2+\bar N^2} \right)
\cong 4j_{k\ell }^2+4j_{k\ell }+1=\left( {2j_{k\ell }+1} \right)^2,
\end{equation}
therefore ${1 \over 2}\left( {2j_{k\ell }+1} \right)=A_{k\ell }$ where $A_{k\ell }$ is the area of the triangle characterized by
the two vectors $g_k$ and $g_{\ell}$ and $j_{k \ell}$ is the spin corresponding to the representation $\rho _{k\ell }$ associated
to the triangle $kl$. Substituting this result in eq. \ref{g8a} we obtain the asymptotic value of the amplitude given by Barrett and
Williams [19]. 

\section{Conclusion.}
Starting from the Barrett-Crane model, we examine  the geometry and quantization of spin networks in Euclidean space $R^4$. We find that  alternative choices are possible. We follow in part [21]  and quantize a simplicial spin  network by attaching to its boundaries the irreps and intertwiners of the group $SO(4,R)$.
The intertwiners usually are  required to be  
invariant under right action. We point out the equal importance of left action.
A large class of models as [17-23] employs right action invariant intertwiners only between  pairs of irreps. 
Invariance in addition under left action can  
be achieved from full irreps by the formation of traces. 
As an alternative quantization, we follow [22] and examine spherical harmonics 
and their right action invariant intertwiners attached to boundaries of the spin network. The Gelfand-Zetlin basis of the irreps of $SO(4,R)$  
is the appropriate tool for the analysis.
Spherical harmonics by  their transformation properties  select only simple representations. Since spherical harmonics live on the coset space $SO(4,R)/SU(2)^c$,  not on the full group space, their 
intertwiners relate simple representations exclusively  to simple representations.
The pairwise right-invariant intertwiners of spherical harmonics in the Gelfand-Zetlin basis become zonal spherical functions. We construct these explicitly and write them in terms of a single group parameter.
Moreover the zonal spherical functions admit a corresponding geometrical interpretation 
in terms of the area of triangles.
In part II we shall develop a similar analysis for relativistic spin networks 
in Minkowski space.

\acknowledgements
One of the author (M.L.) expresses his gratitude to the Director of the Institut f\"{u}r Theoretische Physik, of the University of T\"ubingen, where part of this work was done, and to Prof. Barrett for illuminating conversations about spherical functions. This work has been partially supported by M.E.C.
(Spain). Grant: FPA2006-09199.

\end{article}
\end{document}